\documentclass[12pt, a4paper]{article}
\usepackage{graphicx}

\newcommand{\D}{\mbox{\rm d}}
\newcommand{\T}{\mathcal{T}}
\newcommand{\R}{\mathcal{R}}
\newcommand{\A}{\mathcal{A}}

\begin{document}

\title{\large \bf CHARACTERIZATION OF UNWANTED NOISE\\ IN REALISTIC CAVITIES
\footnote{Contribution to XI International Conference on Quantum
Optics, Minsk, Belarus, 26-31 May, 2006}}
\author{A. A. Semenov$^{1,3}$, D. Yu. Vasylyev$^{1,3}$,
  W. Vogel$^1$,\\
M. Khanbekyan$^2$, and D.-G. Welsch$^2$\\ \quad \\
{\footnotesize $^{1}$Institut f\"ur Physik, Universit\"{a}t Rostock,}\\
{\footnotesize Universit\"{a}tsplatz 3, D-18051 Rostock, Germany}\\
{\footnotesize $^2$Theoretisch-Physikalisches Institut,
Friedrich-Schiller-Universit\"{a}t Jena,}\\{\footnotesize
Max-Wien-Platz 1, D-07743 Jena, Germany}\\
{\footnotesize $^3$Institute of Physics, National Academy of
Sciences of Ukraine,}\\
{\footnotesize Prospect Nauky 46, UA-03028 Kiev, Ukraine}\\
{\footnotesize E-mail: sem@iop.kiev.ua}}

\date{\today}
\maketitle

\begin{abstract}
The problem of the description of absorption and scattering losses
in high-$Q$ cavities is studied. The considerations are based on
quantum noise theories, hence the unwanted noise associated with
scattering and absorption is taken into account
by introduction of additional damping and noise terms in the
quantum Langevin equations and input--output relations.
Completeness conditions for the description of the cavity models
obtained in this way are studied and corresponding
replacement schemes are discussed.
\end{abstract}


\section{Introduction}

Unwanted noise associated with absorption and scattering in high-$Q$
cavities usually plays a crucial role in experiments in cavity
quantum electrodynamics (cavity QED) \cite{CQED}. Even small values
of the corresponding absorption/scattering coefficients may lead to
dramatic changes of the quantum properties of the radiation.
For typical high-$Q$ cavities the unwanted losses can be of the
same order of magnitude as the wanted, radiative losses due to the
input--output coupling \cite{Hood}.
In such a case the process of quantum-state extraction
from a high-$Q$ cavity is characterized by efficiency of about 50\%,
\cite{Khanbekyan}. This feature gives a serious restriction for the
implementation of many proposals in cavity QED. Particularly,
nowadays a lot of schemes for quantum-state engineering of the
intracavity field are known. For example, in Ref.~\cite{Law} a scheme for
the generation of an arbitrary quantum state of the field is
proposed. Also schemes for the generation of entangled states are
known \cite{Lange}. Unfortunately, due to the small efficiency of
the quantum-state extraction, the states of the field may lose
essential nonclassical properties after escaping from the cavity.

In the framework of quantum noise theories (QNT), a high-$Q$-cavity
mode is usually considered as a harmonic oscillator
interacting through the coupling mirror with a number of external
modes. This leads to the description of the cavity mode in terms of
a quantum Langevin equation and input--output relation
\cite{Collett}. The same result can be obtained in a quantum-field
theoretical (QFT) approach \cite{Knoell}-\cite{Viviescas}
in appropriate limits.

For the description of unwanted losses of the intracavity field, it
is sufficient to suppose the existence of a non-radiative
input--output channel associated with absorption and scattering
processes \cite{Khanbekyan, Viviescas}. However this model works
properly, with respect to the output field, only in cases when the
input ports of the cavity are unused. Indeed, it is clear that an
input field can be absorbed or scattered in the coupling mirror. A
detailed analysis shows that for a complete description of the
unwanted noise one should take into account also the
absorption/scattering losses inside the coupling mirrors
\cite{khanbekyan3}.

As we will show below, the unwanted noise can be modeled by
introducing blocks of beam
splitters in an appropriately chosen replacement scheme,
leading to
additional noise terms in the quantum Langevin equation and the
input--output relation. Such a description allows one to have a clear
geometrical interpretation of the operators of unwanted noise as
vectors in a unitary space. The minimum dimension of this space
depends on the number of input--output ports. For example, for a one
sided-cavity a two-dimensional space is sufficient,
two-sided cavities require a three dimensional space and so on.

The requirement of preserving equal-time commutation rules leads to the
conclusion that the $c$-number coefficients in the quantum Langevin
equation and the input--output relations satisfy several constraints.
In other words, their values belong to a certain multidimensional
manifold. Therefore, the problem of consistency and completeness of
the corresponding replacement scheme can be solved by applying
differential geometry \cite{Dubrovin}.

In the present paper we formulate conditions of completeness for
replacement schemes modeling unwanted noise in high-$Q$ cavities and
consider several examples. The paper is organized as follows. In
Sec.~\ref{Completeness} the mathematical model of unwanted noise
in one-sided cavities is introduced. Examples of
replacement schemes are considered in Sec.~\ref{ReplacementSchemes}
and their applicability is discussed. Cavities with
two and more input--output ports are considered in
Sec.~\ref{TwoSidedCavities}. A summary and some concluding remarks
are given in Sec.~\ref{Conclusions}.

\section{Complete model of the unwanted noise}
\label{Completeness}

\subsection{Idealized cavity model}

Let us start our consideration by considering a one-sided cavity and
reminding the standard QNT model,
which does not include channels of unwanted noise \cite{Collett,
Knoell}. In this case the cavity-mode operator
$\hat{a}_\mathrm{cav}$ obeys the quantum Langevin equation
\begin{equation}
\dot{\hat{a}}_\mathrm{cav} (t)=-\left(i\omega_\mathrm{cav}+
\frac{\Gamma}{2}\right)\hat{a}_\mathrm{cav}(t)+
\T^\mathrm{(c)}\,\hat{b}_\mathrm{in}\left(t\right),\label{QLEStandard}
\end{equation}
where $\omega_\mathrm{cav}$ is a resonant frequency of the cavity,
$\Gamma$ is the cavity decay rate, $\T^\mathrm{(c)}$ is the complex
transmission coefficient describing injection of an input field into
the cavity, and $\hat{b}_\mathrm{in}\left(t\right)$ is the
input-field operator satisfying the commutation rule
\begin{equation}
\left[\hat{b}_\mathrm{in}(t_1), \hat{b}_\mathrm{in}^\dag(t_2)
\right]=\delta(t_1-t_2). \label{InCommutator}
\end{equation}
The output-field operator $\hat{b}_\mathrm{out}\left(t\right)$
satisfying the commutation rule
\begin{equation}
\left[\hat{b}_\mathrm{out}(t_1), \hat{b}_\mathrm{out}^\dag(t_2)
\right]=\delta(t_1-t_2) \label{OutCommutator}
\end{equation}
is connected to the cavity-mode operator
$\hat{a}_\mathrm{cav}\left(t\right)$ and the input-field operator
$\hat{b}_\mathrm{in}\left(t\right)$ via the input--output relation
\begin{equation}
\hat{b}_\mathrm{out}\left(t\right)
=\T^\mathrm{(o)}\,\hat{a}_\mathrm{cav}\left(t\right)+\R^\mathrm{(o)}
\hat{b}_\mathrm{in}\left(t\right).\label{IORStandard}
\end{equation}
Here $\T^\mathrm{(o)}$ is the complex transmission coefficient
so that extraction of the cavity field becomes possible
Ref.~\cite{Khanbekyan}, and $\R^\mathrm{(o)}$ is the complex
reflection coefficient at the cavity.

The solution of the quantum Langevin equation (\ref{QLEStandard})
can be written in the form
\begin{equation}
\hat{a}_\mathrm{cav}\!(t) =
\hat{a}_\mathrm{cav}\!(0)e^{-(i\omega_\mathrm{cav}+\Gamma/2)t}
+\T^\mathrm{(c)}\int_{0}^{t}\D t^{\prime}
e^{-(i\omega_\mathrm{cav}+\Gamma/2)(t-t^{\prime})}
\hat{b}_\mathrm{in}\!(t^{\prime}).\label{QLEStandardSolution}
\end{equation}
Inserting it into the input--output relation (\ref{IORStandard}), one
obtains an expression for the output-field operator in terms of the
input-field operator and the operator of the cavity mode at the
initial time,
\setlength\arraycolsep{1pt}
\begin{eqnarray}
\hat{b}_\mathrm{out}\!\left(t\right)
&=&\T^\mathrm{(o)}\hat{a}_\mathrm{cav}\!(0)\,e^{-(i\omega_\mathrm{cav}+\Gamma/2)t}\nonumber\\
&+&\T^\mathrm{(o)}\T^\mathrm{(c)}\int_{0}^{t}\D t^{\prime}
e^{-(i\omega_\mathrm{cav}+\Gamma/2)(t-t^{\prime})}
\hat{b}_\mathrm{in}\!(t^{\prime})+\R^\mathrm{(o)}
\hat{b}_\mathrm{in}\!\left(t\right).\label{IORStandardExtended}
\end{eqnarray}
Assuming that the cavity-mode operator obeys the standard bosonic
commutation relation at all times
\begin{equation}
\left[\hat{a}_\mathrm{cav}(t), \hat{a}_\mathrm{cav}^\dag(t)
\right]=1, \label{CavCommutator}
\end{equation}
and that the input-field operator commutes
with the cavity-mode operator at the initial time $t=0$, using
Eqs.~(\ref{InCommutator}, \ref{QLEStandardSolution},
\ref{CavCommutator}) one obtains the constraint
\begin{equation}
\Gamma=\left|\T^\mathrm{(c)}\right|^2.\label{Constraint1}
\end{equation}
In a similar way from Eqs.~(\ref{InCommutator}, \ref{OutCommutator},
\ref{IORStandardExtended}, \ref{CavCommutator}) we can see that the
conditions
\begin{equation}
\left|\R^\mathrm{(o)}\right|^2=1,\label{Constraint2}
\end{equation}
and
\begin{equation}
\T^\mathrm{(o)}+\T^{\mathrm{(c)}\ast}\R^\mathrm{(o)}=0 \label{Constraint3}
\end{equation}
are satisfied.
A consequence of these constraints is the fact that the reflection
coefficient $\R^\mathrm{(o)}$ can be expressed in terms of
$\T^\mathrm{(o)}$ and $\T^\mathrm{(c)}$ as
\begin{equation}
\R^\mathrm{(o)}=-\frac{\T^\mathrm{(o)}}{\T^\mathrm{(c)\ast}}.\label{RoExpression}
\end{equation}

In summary, the QNT model of a cavity without unwanted noise includes the
quantum Langevin equation (\ref{QLEStandard}), the input--output
relation (\ref{IORStandard}) and the constraints for the $c$-number
coefficients (\ref{Constraint1}-\ref{Constraint3}). In particular, the
constraint (\ref{Constraint1}) describes the relation between the cavity decay
rate and the coefficient $\T^\mathrm{(c)}$. A consequence of
the constraints is the equality of the absolute values of the
transmission coefficients $\T^\mathrm{(o)}$ and $\T^\mathrm{(c)}$.

\subsection{Realistic cavity model}

As it has already been mentioned in the introduction, unwanted noise
can be included in the QNT model through introduction of
additional noise terms. Indeed, for a description of
absorption and scattering of the cavity field, one can
consider the quantum Langevin equation
\begin{equation}
\dot{\hat{a}}_\mathrm{cav}=-\left[i\omega_\mathrm{cav}+ {\textstyle
\frac{1}{2} } \Gamma\right]\hat{a}_\mathrm{cav}
 +\,\T^\mathrm{(c)}\hat{b}_\mathrm{in}\left(t\right)
+\hat{C}^{(c)}\left(t\right). \label{QLENoise}
\end{equation}
Accordingly, the possibility of absorption and scattering
of the input field is described by introduction of an
additional term into the input--output relation
\begin{equation}
\hat{b}_\mathrm{out}\left(t\right)=
\T^\mathrm{(o)}\hat{a}_\mathrm{cav}\left(t\right)
+\R^\mathrm{(o)}\hat{b}_\mathrm{in}\left(t\right)
+\hat{C}^{(o)}\left(t\right). \label{IORNoise}
\end{equation}
In these equations, the operators of unwanted noise,
$\hat{C}^{(c)}\left(t\right)$ and $\hat{C}^{(o)}\left(t\right)$,
which commute with the input-field operator $\hat{b}_\mathrm{in}\left(t\right)$,
the cavity-field operator at the initial time $t=0$,
$\hat{a}_\mathrm{cav}\left(0\right)$, satisfy the following
commutation rules:
\begin{equation}
\left[\hat{C}^{(c)}(t_1),\hat{C}^{(c)\dag}(t_2)\right]=
\left|\A^{(c)}\right|^2\delta(t_1-t_2),\label{CommutatorC1}
\end{equation}
\begin{equation}
\left[\hat{C}^{(o)}(t_1),\hat{C}^{(o)\dag}(t_2)\right]=
\left|\A^{(o)}\right|^2\delta(t_1-t_2),\label{CommutatorC2}
\end{equation}
\begin{equation}
\left[\hat{C}^{(c)}(t_1),\hat{C}^{(o)\dag}(t_2)\right]=
\Xi\,\delta(t_1-t_2).\label{CommutatorC3}
\end{equation}
Here, $\A^{(c)}$, $\A^{(o)}$, and $\Xi$, are $c$-number
absorption/scattering coefficients. The set of the coefficients
$\Gamma$, $\omega_\mathrm{cav}$, $\T^{(c)}$, $\T^{(o)}$, $\R^{(o)}$,
$\left|\A^{(c)}\right|^2$, $\left|\A^{(o)}\right|^2$, and $\Xi$
characterizes the cavity with unwanted noise.

Similar to the case of an idealized cavity, one can show that the
requirement of preserving the commutation rules leads to the
following constraints:
\begin{equation}
\Gamma=\left|{\A}^\mathrm{(c)}\right|^2+
\left|\T^\mathrm{(c)}\right|^2, \label{ConstraintNoise1}
\end{equation}
\begin{equation}
\left|\R^\mathrm{(o)}\right|^2+\left|\A^\mathrm{(o)}\right|^2=1,
\label{ConstraintNoise2}
\end{equation}
\begin{equation}
\T^\mathrm{(o)}+\T^{\mathrm{(c)}\ast}\R^\mathrm{(o)}+
\Xi=0.\label{ConstraintNoise3}
\end{equation}
Hence, we conclude that the $c$-number coefficients describing a
realistic cavity should belong to the manifold defined by
Eqs.~(\ref{ConstraintNoise1}-\ref{ConstraintNoise3}). As it is
well known from differential geometry, each manifold can be
described by means of independent parameters \cite{Dubrovin}.
Particularly, this means that the $c$-number coefficients can be
expressed in terms of appropriately chosen parameters. The corresponding
parametrization should cover the whole manifold. In other words, it
should describe all possible cavities. Otherwise, one gets a
degenerate model, which describes just a restricted class of the
cavities.

The easiest way for the parametrization of the considered manifold
follows directly from
Eqs.~(\ref{ConstraintNoise1}-\ref{ConstraintNoise3}). Indeed, the
coefficients $\Gamma$, $\omega_\mathrm{cav}$, $\T^{(c)}$,
$\T^{(o)}$, $\R^{(o)}$ can be considered as independent parameters,
and the coefficients $\left|\A^{(c)}\right|^2$,
$\left|\A^{(o)}\right|^2$, $\Xi$ are expressed in terms of them.
Such a parametrization can be convenient in experimental
investigations, because the corresponding parameters have a clear
physical interpretation. However, in the theoretical investigation
one should use this parametrization very carefully. In fact, not all
values of these independent parameters belong to the manifold.
Particularly, their values are necessarily restricted by the
inequalities
\begin{equation}
\left|\T^\mathrm{(c)}\right|^2\leq\Gamma, \label{Ineq1}
\end{equation}
\begin{equation}
\left|\T^\mathrm{(o)}\right|^2\leq\Gamma, \label{Ineq2}
\end{equation}
\begin{equation}
\left|\R^\mathrm{(o)}\right|^2\leq 1, \label{Ineq3}
\end{equation}
\begin{equation}
\left|\T^\mathrm{(o)}+\T^\mathrm{(c)*}\R^\mathrm{(o)}\right|^2\leq
\left(
\Gamma-\left|\T^\mathrm{(o)}\right|^2\right)\left(1-\left|\R^\mathrm{(o)}
\right|^2\right), \label{Ineq4}
\end{equation}
which are derived from
Eqs.~(\ref{ConstraintNoise1}-\ref{ConstraintNoise3}). The fact that
the parameters satisfy these inequalities is not a sufficient
condition for describing realistic cavities. However, if they do not
satisfy these conditions, one can conclude that the chosen values
are not physically consistent ones.

\subsection{Operators of unwanted noise}

It is worth noting that the operators of unwanted noise,
$\hat{C}^{(c)}(t_1 )$ and $\hat{C}^{(o)}(t_1 )$, can be considered
as two vectors in a unitary vector space. In particular, the scalar
product of two arbitrary vectors $\hat{C}^{(a)} (t)$ and
$\hat{C}^{(b)} (t)$ in this space can be defined by
\begin{equation}
\left(\hat{C}^{(a)},\hat{C}^{(b)}\right) =\int\D t_1\,
\left[\hat{C}^{(a)\dag}(t_1),
\hat{C}^{(b)}(t_2)\right].\label{ScalarProduct}
\end{equation}
In this interpretation, $|{\A}^\mathrm{(c)}|$ and
$|{\A}^\mathrm{(o)}|$ can be considered as the absolute values of
the vectors $\hat{C}^{(c)}(t)$ and $\hat{C}^{(o)}(t)$, respectively,
whereas $\Xi$ defines the (complex) angle between them, i.e.,
\begin{equation}
\Xi=\left|\A^{(c)}\right|\left|\A^{(o)}\right|e^{i\kappa}\cos\zeta,
\label{XiProperty}
\end{equation}
where $\kappa$ and $\zeta$ are real.

\begin{figure}[ht!]
\begin{center}
\includegraphics[width=0.5\linewidth,
clip=]{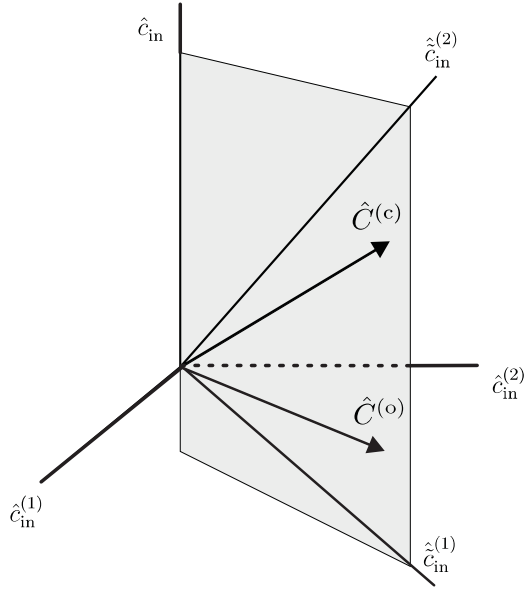} \caption{\label{FigGeom} Geometrical interpretation
of the operators of the unwanted noise. The operators
$\hat{{c}}^{(1)}_\mathrm{in}(t)$, $\hat{{c}}^{(2)}_\mathrm{in}(t)$
and $\hat{c}_\mathrm{in}(t)$ correspond to a three-dimensional
representation of the unwanted noise. The operators
$\hat{\tilde{c}}^{(1)}_\mathrm{in}(t)$ and
$\hat{\tilde{c}}^{(2)}_\mathrm{in}(t)$ correspond to an equivalent
representation in a two-dimensional space. }
\end{center}
\end{figure}

The vectors $\hat{C}^{(c)}(t)$ and $\hat{C}^{(o)}(t)$ can be
expanded in an orthogonal basis,
\begin{eqnarray}
\hat{C}^{(c)}(t)=\sum_k
{\A}^\mathrm{(c)}_{(k)}\hat{{c}}^{(k)}_\mathrm{in}(t),
\label{BasisC}\\
\hat{C}^{(o)}(t) =\sum_k
{\A}^\mathrm{(o)}_{(k)}\hat{{c}}^{(k)}_\mathrm{in}(t) ,
\label{BasisO}
\end{eqnarray}
which implies that different representations of the operators of the
unwanted noise can be obtained. It is clear (in full analogy with
usual geometry) that two vectors always belong to a two-dimensional
plane, see Fig.~\ref{FigGeom}. This plane may be considered as a
two-dimensional unitary vector space, with the two basis vectors
$\hat{\tilde{c}}^{(1)}_\mathrm{in}(t)$ and
$\hat{\tilde{c}}^{(2)}_\mathrm{in}(t)$ playing the role of
appropriately chosen operators of the unwanted noise. Consequently,
it is sufficient to have two basis operators for a complete
description of the unwanted noise of a (one-sided) cavity. However,
as we will show in the following, in some cases it is convenient to
use representations with a lager number of dimensions, with
particular emphasis on the three-dimensional case.

\section{Replacement schemes}
\label{ReplacementSchemes}

\subsection{Complete scheme}

\begin{figure}[ht!]
\begin{center}
\includegraphics[width=0.75\linewidth,
clip=]{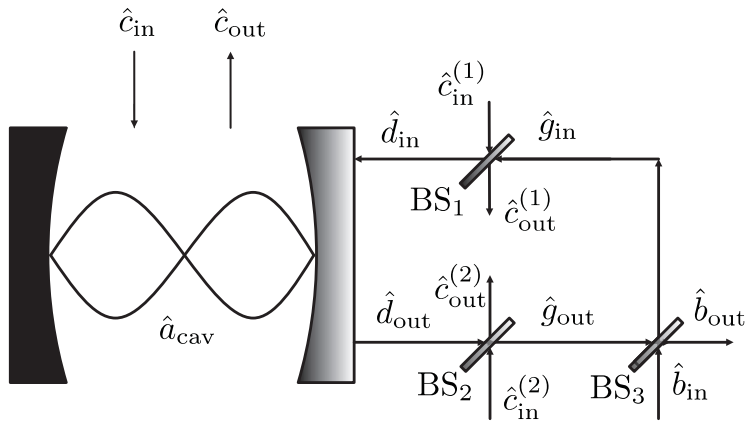} \caption{\label{ReplScheme} Replacement scheme for
the simulation of unwanted noise in a high-$Q$ cavity. The $SU(2)$
beam splitters $\mathrm{BS}_1$ and $\mathrm{BS}_2$ model unwanted
noise inside the coupling mirror and $U(2)$ beam splitter
$\mathrm{BS}_3$ simulates feedback.}
\end{center}
\end{figure}

The method of replacement schemes, widely used in quantum optics,
can be applied to the formulation of a parametrization which works
with all the values of independent parameters. The main idea is to
simulate the channels of unwanted noise by an additional input--output
port and a block of beam splitters as illustrated in Fig.~\ref{ReplScheme}.
The additional
input--output port models the losses typically responsible for
the absorption and scattering of the radiation, which escapes from the
cavity, whereas the block of beam splitters describes losses like
absorption and scattering of the input field entering
the cavity. In Fig.~\ref{ReplScheme},
the symmetrical beam-splitters $\mathrm{BS}_1$ and $\mathrm{BS}_2$
simulate the unwanted noise inside the coupling mirror. Moreover, due to
the requirements of completeness one must include in the scheme
the asymmetrical $U(2)$ beam-splitter $\mathrm{BS}_3$, which
simulates feedback.

Using the quantum Langevin equation and the input--output relation
for a cavity with two input--output ports as well as the
input--output relations for each beam splitter separately, we obtain
Eqs.~(\ref{IORStandard}, \ref{QLENoise}), where the operators
$\hat{C}^{(c)}(t)$ and $\hat{C}^{(o)}(t)$ have the form
\begin{equation}
\hat{C}^{(c)}(t)=\A^\mathrm{(c)}_{(1)}\hat{c}^{(1)}_\mathrm{in}\left(t\right)
+\A^\mathrm{(c)}_{(2)}\hat{c}^{(2)}_\mathrm{in}\left(t\right)
+\A\hat{c}_\mathrm{in}\left(t\right),
\label{ReplC}
\end{equation}
\begin{equation}
\hat{C}^{(o)}(t)
=\A^\mathrm{(o)}_{(1)}\hat{c}^{(1)}_\mathrm{in}\left(t\right)
+\A^\mathrm{(o)}_{(2)}\hat{c}^{(2)}_\mathrm{in}\left(t\right).
\label{ReplO}
\end{equation}
The $c$-number coefficients in the quantum Langevin equation and the
input--output relation are expressed in terms of the beam-splitter
transmission and reflection coefficients $\T^{(k)}$ and $\R^{(k)}$
respectively ($k=1,2,3$), the phase factor $\varphi^{(3)}$ of the
beam splitter $\mathrm{BS}_3$, the resonance frequency $\omega_0$,
the radiation and absorption decay rates of the ``primary'' cavity
in the scheme, $\gamma$ and $\left|\A\right|^2$ respectively, as
follows:
\begin{equation}
\Gamma=\gamma\,\frac{1-\left|\R^{(3)}\right|^2\left|\T^{(1)}\right|^2
\left|\T^{(2)}\right|^2}
{\left|1-\R^{(3)\ast}\T^{(1)}\T^{(2)}\right|^2} +\left|\A\right|^2
\label{Gamma}
\end{equation}
\begin{equation}
\omega_\mathrm{cav} =\omega_{0} - i\frac {\gamma} {2}
\frac{\R^{(3)\ast}\T^{(1)}\T^{(2)}-
\R^{(3)}\T^{(1)\ast}\T^{(2)\ast}}
{\left|1-\R^{(3)\ast}\T^{(1)}\T^{(2)}\right|^2} \label{Omega}
\end{equation}
\begin{equation}
\T^\mathrm{(c)}=\sqrt{\gamma}\,\frac{\T^{(1)}\T^{(3)\ast}}
{1-\R^{(3)\ast}\T^{(1)}\T^{(2)}}\label{Tc}
\end{equation}
\begin{equation}
\A^\mathrm{(c)}_{(1)}=\sqrt{\gamma}\,\frac{\R^{(1)}}
{1-\R^{(3)\ast}\T^{(1)}\T^{(2)}},\label{Ac1}
\end{equation}
\begin{equation}
\A^\mathrm{(c)}_{(2)}=-\sqrt{\gamma}\,\frac{\T^{(1)}\R^{(2)}\R^{(3)\ast}}
{1-\R^{(3)\ast}\T^{(1)}\T^{(2)}},\label{Ac2}
\end{equation}
\begin{equation}
\T^\mathrm{(o)}=\sqrt{\gamma}\,e^{i\varphi^{(3)}}\frac{\T^{(2)}\T^{(3)}}
{1-\R^{(3)\ast}\T^{(1)}\T^{(2)}},\label{To}
\end{equation}
\begin{equation}
\R^\mathrm{(o)}=e^{i\varphi^{(3)}}\frac{\R^{(3)}-\T^{(1)}\T^{(2)}}
{1-\R^{(3)\ast}\T^{(1)}\T^{(2)}},\label{Ro}
\end{equation}
\begin{equation}
\A^\mathrm{(o)}_{(1)}=-e^{i\varphi^{(3)}}\frac{\T^{(2)}\R^{(1)}\T^{(3)}}
{1-\R^{(3)\ast}\T^{(1)}\T^{(2)}},\label{Ao1}
\end{equation}
\begin{equation}
\A^\mathrm{(o)}_{(2)}=e^{i\varphi^{(3)}}\frac{\R^{(2)}\T^{(3)}}
{1-\R^{(3)\ast}\T^{(1)}\T^{(2)}}.\label{Ao}
\end{equation}
Each complex coefficient $\T^{(k)}$ and $\R^{(k)}$ can be expressed
in terms of three real independent parameters $\theta^{(k)}$,
$\mu^{(k)}$ and $\nu^{(k)}$:
\begin{eqnarray}
\T^{(k)}=\cos{\theta^{(k)}}e^{i\mu^{(k)}},\label{Tk}\\
\R^{(k)}=\sin{\theta^{(k)}}e^{i\nu^{(k)}}.\label{Rk}
\end{eqnarray}
Therefore, one gets the parametrization of the manifold by means of
the independent parameters.

In order to check the completeness of the proposed parametrization,
one should first present Eqs.~(\ref{Gamma}-\ref{Ao}) in the form of
real functions of real parameters. Next, one should build the matrix
containing the first derivatives of these functions. The determinant
of this function should be non-zero. The corresponding calculations
have been performed by using Mathematica 5.1. It has been found that
the parametrization corresponding to the considered replacement
scheme and given by Eqs.~(\ref{Gamma}-\ref{Ao}) completely describes
(one-sided) cavities with unwanted noise.

\subsection{Degenerate schemes}

Let us consider examples of replacement schemes, referred to
as degenerate replacement schemes,
which do not describe all possible cavities.
As a rule, degenerate schemes can be obtained
from the complete scheme by removing one or more elements. Cavities
modeled by degenerate replacement schemes usually obey some
constraints in addition to
Eqs.~(\ref{ConstraintNoise1}-\ref{ConstraintNoise3}). In other
words, such cavities correspond to points in a certain sub-manifold
rather than in the whole manifold.

The first example is the class of cavities obtained from
Fig.~\ref{ReplScheme} by removing the beam splitters $\mathrm{BS}_1$
and $\mathrm{BS}_2$. For such cavities, the input field does
not suffer from losses when it enters the cavity. The operators
$\hat{C}^{(c)}(t)$ and $\hat{C}^{(o)}(t)$ in the quantum Langevin
equation and the input--output relation in this case read as
\begin{equation}
\hat{C}^{(c)}(t)=\A\,\hat{c}_\mathrm{in}\left(t\right),
\label{ReplCDeg1}
\end{equation}
\begin{equation}
\hat{C}^{(o)}(t) =0. \label{ReplODeg1}
\end{equation}
It is clear that for such a cavity the noise term associated with
absorption and scattering is only included in the quantum Langevin
equation -- a model, which can be used for special applications
\cite{Khanbekyan, Viviescas}. The corresponding
parametrization is a rather trivial one, the additional constraint
has the form
\begin{equation}
\left|\R^\mathrm{(o)}\right|^2=1.\label{ConstraintDeg1}
\end{equation}
A consequence of this fact is that $\A^\mathrm{(o)}=0$. Moreover,
such a cavity has some properties very close to the idealized cavity
without channels of unwanted noise. The transmission coefficients
$\T^\mathrm{(o)}$ and $\T^\mathrm{(c)}$ are equal and
Eq.~(\ref{RoExpression}) holds true for the reflection coefficient
$\R^\mathrm{(o)}$.

Another example of a degenerate scheme can be obtained from the
complete scheme in Fig.~\ref{ReplScheme} by removing the (non-radiative)
input--output channels $\hat{c}_\mathrm{in}$, $\hat{c}_\mathrm{out}$
and the beam splitter $\mathrm{BS}_3$ associated with the feedback.
In this case, the operators $\hat{C}^{(c)}(t)$ and $\hat{C}^{(o)}(t)$
have the form
\begin{equation}
\hat{C}^{(c)}(t)={\A}^\mathrm{(c)}_{(1)}\,\hat{{c}}^{(1)}_\mathrm{in}(t),
\label{ReplCDeg2}
\end{equation}
\begin{equation}
\hat{C}^{(o)}(t)
=\A^\mathrm{(o)}_{(1)}\,\hat{c}^{(1)}_\mathrm{in}(t)
+\A^\mathrm{(o)}_{(2)}\,\hat{c}^{(2)}_\mathrm{in}(t).\label{ReplODeg2}
\end{equation}
Although both the quantum Langevin equation and the
input--output relation contain noise terms associated with
unwanted losses, the scheme is not a complete one.
The corresponding parametrization can be written in the form
\begin{equation}
\Gamma=\gamma, \label{GammaD}
\end{equation}
\begin{equation}
\omega_\mathrm{cav}=\omega_{0} , \label{OmegaD}
\end{equation}
\begin{equation}
\T^\mathrm{(c)}=\sqrt{\gamma}\,\T^{(1)}, \label{TcD}
\end{equation}
\begin{equation}
{\A}^\mathrm{(c)}_{(1)}=\sqrt{\gamma}\,\R^{(1)}, \label{AcD}
\end{equation}
\begin{equation}
\T^\mathrm{(o)}=\sqrt{\gamma}\,\T^{(2)}, \label{ToD}
\end{equation}
\begin{equation}
\R^\mathrm{(o)}=-\T^{(1)}\T^{(2)}, \label{RoD}
\end{equation}
\begin{equation}
{\A}^\mathrm{(o)}_{(1)}=-\R^{(1)}\T^{(2)}, \label{Ao1D}
\end{equation}
\begin{equation}
{\A}^\mathrm{(o)}_{(2)}=\R^{(2)}. \label{Ao2D}
\end{equation}
One can easily prove by direct calculations that along with
Eqs.~(\ref{ConstraintNoise1}-\ref{ConstraintNoise3}) this
parametrization satisfy the following additional constraint
\begin{equation}
\frac{\T^\mathrm{(o)}\T^\mathrm{(c)}}{\Gamma}+\R^\mathrm{(o)}=0.
\label{ConstraintDeg2}
\end{equation}
It is worth noting that the physics behind this degenerate scheme is
closely related to that of a cavity without unwanted noise. Indeed,
the unwanted
noise can be regarded as noise associated with the transmission
channel. Hence, the losses modeled in this way cannot affect the decay
rate of the intracavity field,
but some properties of the external field are changed.
Particularly, for such cavities it is impossible to combine
a cavity mode and an input mode in an output mode.

\section{Two-sided cavities}
\label{TwoSidedCavities}

\begin{figure}[ht!]
\begin{center}
\includegraphics[width=0.95\linewidth,clip=]{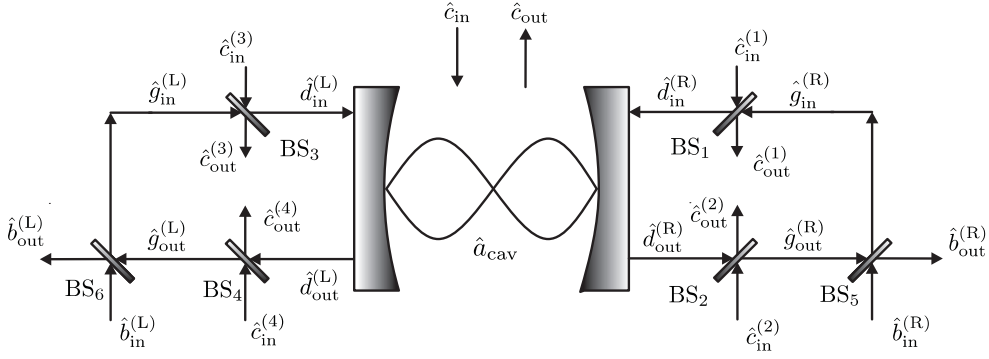}
\caption{\label{figtwosides} Replacement scheme for modeling
unwanted noise in a two-sided cavity. The symmetrical $SU(2)$-type
beam splitters $\mathrm{BS}_1$, $\mathrm{BS}_2$, $\mathrm{BS}_3$, and
$\mathrm{BS}_4$ model the unwanted noise in the two coupling
mirrors, and the asymmetrical $U(2)$-type beam splitters
$\mathrm{BS}_5$ and $\mathrm{BS}_6$ simulate some feedback.}
\end{center}
\end{figure}%

So far we have considered one-sided cavities.
In various physical applications, e.g., the generation
of squeezed light in optical parametric amplification \cite{Wu},
it is necessary to consider the problem of unwanted noise in
cavities with two (or even more) radiative
input--output ports.
Let us generalize the replacement-scheme method developed
above to a two-sided cavity, as is sketched
in Fig.~\ref{figtwosides}. It is straightforward to show
that the generalization of Eqs.~(\ref{QLENoise}) and
(\ref{IORNoise}) is
\begin{eqnarray}
\dot{\hat{a}}_\mathrm{cav}&=&-\left[i\omega_\mathrm{cav}
 +
{\textstyle \frac{1}{2}} \Gamma\right]\hat{a}_\mathrm{cav}
 +\T^\mathrm{(c)}_\mathrm{(R)}\hat{b}^\mathrm{(R)}_\mathrm{in}(t)
 +\T^\mathrm{(c)}_\mathrm{(L)}\hat{b}^\mathrm{(L)}_\mathrm{in}(t)
\nonumber\\&+&\A^\mathrm{(c)}_{(1)}\hat{c}^{(1)}_\mathrm{in}(t)
+\A^\mathrm{(c)}_{(2)}\hat{c}^{(2)}_\mathrm{in}(t)
+\A^\mathrm{(c)}_{(3)}\hat{c}^{(3)}_\mathrm{in}(t)
+\A^\mathrm{(c)}_{(4)}\hat{c}^{(4)}_\mathrm{in}(t)
+\A\hat{c}_\mathrm{in}(t),\label{ts5}
\end{eqnarray}
\begin{eqnarray}
\hat{b}^\mathrm{(R)}_\mathrm{out}(t)
=\T^\mathrm{(o)}_\mathrm{(R)}\hat{a}_\mathrm{cav}(t)
+\R^\mathrm{(o)}_\mathrm{(R)}\hat{b}^\mathrm{(R)}_\mathrm{in}(t)
\nonumber\\+\, \A^\mathrm{(o)}_{(1)}\hat{c}^{(1)}_\mathrm{in}(t)
+\A^\mathrm{(o)}_{(2)}\hat{c}^{(2)}_\mathrm{in}(t),\label{ts6}
\end{eqnarray}
\begin{eqnarray}
\hat{b}^\mathrm{(L)}_\mathrm{out}(t)
=\T^\mathrm{(o)}_\mathrm{(L)}\hat{a}_\mathrm{cav}(t)
+\R^\mathrm{(o)}_\mathrm{(L)}\hat{b}^\mathrm{(L)}_\mathrm{in}(t)
\nonumber\\+\, \A^\mathrm{(o)}_{(3)}\hat{c}^{(3)}_\mathrm{in}(t)
+\A^\mathrm{(o)}_{(4)}\hat{c}^{(4)}_\mathrm{in}(t),\label{ts7}
\end{eqnarray}
where the coefficients in these equations can be obtained in a
similar manner as for the case of a one-sided cavity.

As in the case of a one-sided cavity, the $c$-number coefficients in
Eqs.~(\ref{ts5})--(\ref{ts7}) are also not independent of each
other. From considering the commutation relation for the
cavity-mode operator, one obtains
\begin{eqnarray}
\Gamma= \bigl|\A\bigr|^2+\bigl|\A^\mathrm{(c)}_{(1)}\bigr|^2+
\bigl|\A^\mathrm{(c)}_{(2)}\bigr|^2+
\bigl|\A^\mathrm{(c)}_{(3)}\bigr|^2+
\bigl|\A^\mathrm{(c)}_{(4)}\bigr|^2 \nonumber\\
+\bigl|\T^\mathrm{(c)}_\mathrm{(R)}\bigr|^2+
\bigl|\T^\mathrm{(c)}_\mathrm{(L)}\bigr|^2.\label{ts8}
\end{eqnarray}
With regard to the right-hand wall of the cavity, the requirement of
preserving the commutation rules implies that
\begin{equation}
\bigl|\R^\mathrm{(o)}_\mathrm{(R)}\bigr|^2
+\bigl|\A^\mathrm{(o)}_{(1)}\bigr|^2+
\bigl|\A^\mathrm{(o)}_{(2)}\bigr|^2=1, \label{ts9}
\end{equation}
\begin{equation}
\T^\mathrm{(o)}_\mathrm{(R)}
+\T^{\mathrm{(c)}\ast}_\mathrm{(R)}\R^\mathrm{(o)}_\mathrm{(R)}
+\A^{\mathrm{(c)}\ast}_{(1)}\A^\mathrm{(o)}_{(1)}
+\A^{\mathrm{(c)}\ast}_{(2)}\A^\mathrm{(o)}_{(2)}=0.\label{ts10}
\end{equation}
Finally, for the field outgoing from the left-hand wall of the
cavity, one can show that
\begin{equation}
\bigl|\R^\mathrm{(o)}_\mathrm{(L)}\bigr|^2\
+\bigl|\A^\mathrm{(o)}_{(3)}\bigr|^2+
\bigl|\A^\mathrm{(o)}_{(4)}\bigr|^2=1, \label{ts11}
\end{equation}
\begin{equation}
\T^\mathrm{(o)}_\mathrm{(L)}
+\T^{\mathrm{(c)}\ast}_\mathrm{(L)}\R^\mathrm{(o)}_\mathrm{(L)}
+\A^{\mathrm{(c)}\ast}_{(3)}\A^\mathrm{(o)}_{(3)}
+\A^{\mathrm{(c)}\ast}_{(4)}\A^\mathrm{(o)}_{(4)}=0.\label{ts12}
\end{equation}
It should be pointed out that the operators of unwanted
noise -- represented by $\hat{c}_\mathrm{in}$,
$\hat{c}^{(1)}_\mathrm{in}$, $\hat{c}^{(2)}_\mathrm{in}$,
$\hat{c}^{(3)}_\mathrm{in}$ and $\hat{c}^{(4)}_\mathrm{in}$
in Eqs. (\ref{ts5}-\ref{ts7}) -- can be
also represented in other forms. Since the corresponding
expressions in Eqs.~(\ref{ts5})--(\ref{ts7}) contain only three
linear combinations of the operators $\hat{c}_\mathrm{in}$,
$\hat{c}^{(1)}_\mathrm{in}$, $\hat{c}^{(2)}_\mathrm{in}$,
$\hat{c}^{(3)}_\mathrm{in}$ and $\hat{c}^{(4)}_\mathrm{in}$, one can
conclude that there exist equivalent formulations of these equations
with three independent operators of unwanted noise.

\section{Summary and conclusions}
\label{Conclusions}

The concept of replacement schemes is a very helpful tool to study, within
the framework of QNT, the effect of unwanted noise associated
with absorption and scattering in realistic high-$Q$ cavities,
which leads to the appearance of additional noise terms in both the
standard quantum Langevin equations and the standard
input--output relations attributed to them.

An important mathematical feature is
the fact that the $c$-number coefficients in the quantum Langevin
equations and in the input--output relation are not independent ones.
In particular, the requirement of preserving typical commutation
rules leads to the appearance of several constraints. Hence, the
corresponding values of the coefficients can be regarded as belonging
to a certain manifold.

A consistent physical description of realistic cavities requires
to formulate the theory in terms of independent parameters only. In
other words, one must consider the parametrization of the manifold.
So, one can formally express some of the $c$-number coefficients in
terms of the other ones and simply consider the latter as independent
parameters. Another, more physical way is a parametrization on the
basis of appropriately chosen replacement schemes.

The method of replacement schemes in fact allows one to distinguish,
with respect to the unwanted noise, between qualitatively different cavity
models. Roughly speaking, one can distinguish between non-degenerate
and degenerate replacement schemes. In contrast to non-degenerate
schemes, where the parametrization completely describes cavities
with unwanted losses, degenerate schemes do not describe
all possible cavities but only special classes.

\subsection*{Acknowledgement}
This work was supported by Deutsche Forschungsgemeinschaft. A.A.S.
also thanks the President of Ukraine for a research stipend.

\end{document}